\documentclass[twocolumn]{aastex631}
\usepackage{natbib}
\usepackage{amsmath}
\usepackage{graphicx}
\usepackage{xcolor}
\usepackage{hyperref}
\usepackage{longtable}

\shorttitle{Ro-vibrational Spectroscopy of CI Tau}
\shortauthors{Kozdon et al.}
\graphicspath{{./}{figures/}}

\begin{document}
\title{Ro-vibrational Spectroscopy of CI Tau ---\\ Evidence of a Multi-Component Eccentric Disk Induced by a Planet}

\author{Janus Kozdon}
\affiliation{Department of Physics and Astronomy, 118 Kinard Laboratory, Clemson University, Clemson, SC 29634-0978, USA}
\author{Sean D. Brittain}
\affiliation{Department of Physics and Astronomy, 118 Kinard Laboratory, Clemson University, Clemson, SC 29634-0978, USA}
\author{Jeffrey Fung}
\affiliation{Department of Physics and Astronomy, 118 Kinard Laboratory, Clemson University, Clemson, SC 29634-0978, USA}
\author{Josh Kern}
\affiliation{Department of Physics and Astronomy, 118 Kinard Laboratory, Clemson University, Clemson, SC 29634-0978, USA}
\author{Stanley Jensen}
\affiliation{Department of Physics and Astronomy, 118 Kinard Laboratory, Clemson University, Clemson, SC 29634-0978, USA}
\author{John S. Carr}
\affiliation{Department of Astronomy, University of Maryland, College Park, MD 20742, USA}
\author{Joan R. Najita}
\affiliation{NSF's NOIRLab, 950 North Cherry Avenue, Tucson, AZ 85719, USA}
\author{Andrea Banzatti}
\affiliation{Department of Physics, Texas State University, San Marcos, TX 78666, USA}

\begin{abstract}
CI Tau is currently the only T Tauri star with an inner protoplanetary disk that hosts a planet, CI Tau b, that has been detected by a radial velocity survey. This provides the unique opportunity to study disk features that were imprinted by that planet. We present multi-epoch spectroscopic data, taken with NASA IRTF in 2022, of the ${}^{12}$CO and hydrogen Pf$\beta$ line emissions spanning 9 consecutive nights, which is the proposed orbital period of CI Tau b. We find that the star's accretion rate varied according to that 9~d period, indicative of companion driven accretion. Analysis of the ${}^{12}$CO emission lines reveals that the disk can be described with an inner and outer component spanning orbital radii 0.05-0.13~au and 0.15-1.5~au, respectively. Both components have eccentricities of about 0.05 and arguments of periapses that are oppositely aligned. We present a proof-of-concept hydrodynamic simulation that shows a massive companion on a similarly eccentric orbit can recreate a similar disk structure. Our results allude to such a companion being located around an orbital distance of 0.14~au. However, this planet's orbital parameters may be inconsistent with those of CI Tau b whose high eccentricity is likely not compatible with the low disk eccentricities inferred by our model.
\end{abstract}


\section{Introduction} \label{sec:i`ntro}
The protoplanetary disks of T Tauri stars often host substructures such as rings/gaps \citep[e.g.,][]{rings_n_gaps}, spiral arms \citep[e.g.,][]{spirals}, and large-scale asymmetries \citep[e.g.,][]{outflow_jets} (see \cite{PP7_disk_struct} for a review). These features may be produced by processes such as condensation fronts or snow lines \citep[e.g.,][]{Pinilla2017}, radiation pressure \citep[e.g.,][]{Bi2022}, gravitational instabilities \citep[e.g.,][]{Dong2018}, and dynamical interactions with planets \citep[][and references therein]{Paard2022}. Amongst these possibilities, planet-disk interactions have generated the most interest because of the connections with exoplanet discoveries. \par

Exoplanets are now known to be ubiquitous (e.g., \citealt{2014_populations}, \citealt{vdm2021} \& \citealt{Christiansen2022}). The majority of detected exoplanets were found by transit and radial velocity surveys that target mature stars well beyond the stage when gas-rich, planet-building disks are present. On the other hand, detections of planets accompanying T Tauri Stars (TTSs) and other Young Stellar Objects (YSOs) that still have their protoplanetary disks are quite rare. This disparity arises because YSOs exhibit substantial stellar activity that, alongside an optically-thick disk, obscures and dampens planetary signatures. Techniques that characterize planets through their imprints on the protoplanetary disks can supplement traditional planet detection techniques. \par

PDS 70 b \& c are two known protoplanets around a TTS \citep{Keppler2018,Muller2018,Haffert2019}, while AB Aur b \citep{Currie2022} and CI Tau b \citep{Johns2016,2018ApJ...866L...6C} are two of the next likely candidates. Amongst these systems, CI Tau b is unique because it is a potential hot Jupiter that has been detected by Doppler monitoring whereas PDS 70 b \& c and AB Aur b were detected by direct imaging at large separations. This allows for CI Tau b to have a well-constrained mass and orbital parameters \citep{Johns2016,2019ApJ...878L..37F}. As such, one should be able to apply theories of planet-disk interactions to observations of the inner disk of CI Tau, and directly test the theoretical predictions. For example, simulations indicate that a massive companion should induce an eccentricity in the disk \citep{Kley2006,teyssandier17} and that the disk drives an eccentricity on the planet \citep{papaloizou01,2017MNRAS.464L.114R,Duffell15,Ragusa18,Muley19}.

CI Tau is a $\sim$2~Myr old TTS of spectral type K5.5 located at a distance of $d$ = 160$\pm$10~pc \citep{Gaia} with a mass of $M_{*}$ = 1.02$\pm{0.001}$~$M_{\sun}$ \citep{Law2022}. Its hot Jupiter companion, CI Tau b \citep{Johns2016}, has a mass of 11.6$\pm2.8 ~M_{\rm J}$ \citep{2019ApJ...878L..37F} and an eccentricity of $e$ = 0.25$\pm$0.16 \citep{2019ApJ...878L..37F}. CI Tau b's orbital period of $P_{\rm orb}$ = 9$\pm$0.5~d \citep{biddle2021} translates to a semi-major axis of $a$ = 0.085~au. The system parameters utilized in our equations are presented in Table \ref{tab:star}. \par

It is expected for massive companions to carve out deep and wide gaps which are identifiable as a lack of NIR emission in the system's Spectral Energy Distribution (SED) \citep{GAP}. This is not the case for CI Tau as it has a substantial NIR excess \citep{McClure2013}. However, \cite{2021ApJ...921L..34M} found that a gap carved by a companion like CI Tau b can potentially be indistinguishable from an undepleted disk --- at least at the NIR end of the SED. The planet's existence was directly questioned by \cite{2020MNRAS.491.5660D} who suggested that the signatures that have been attributed to the planet can, instead, be replicated by stellar activity. However, \cite{2018ApJ...853L..34B} found separate signatures for the planet's orbital period and the star's rotational period. They argued that both signatures cannot be fully attributed to the star alone. \par

Massive companions may, to some degree, mediate stellar accretion \citep{ORBIT}. \cite{2020MNRAS.495.3920T} found that such modulation can even be driven to the point of matching the companion's orbital period (see also \citealt{1996ApJ...467L..77A, 2016ApJ...827...43M}). Such periodic accretion has been observed in binary systems before, e.g. in photometric monitoring of the young stars DQ Tau \citep{tofflemire2017accretion} and TWA 3A \citep{tofflemire2017pulsed}. We seek to capture similar behavior in the Pf$\beta$ line, which has been calibrated against accretion luminosity \citep{2013ApJ...769...21S}, originating from CI Tau. \par

We utilize high resolution spectroscopy for detailed analysis of the emission line profiles originating from CI Tau to characterize its inner protoplanetary disk. Specifically, we study the ro-vibrational NIR emission of the ${}^{12}$CO; which CTTS systems like CI Tau commonly emits \citep{2003ApJ...589..931N}. The distribution of the emitting gas can be determined from the line profiles given a known stellar mass and disk inclination. Asymmetric features can elude to the presence of disk sub-structures such as disk winds \citep{2011_diskwinds}, circumplanetary disks (CPDs; \cite{HD100546_disappears}) and disk eccentricities \citep{Liskowsky2012}. Also, the disk's atmospheric temperature can be estimated from the relative transition strengths.

The following section (Sec. \ref{sec:obs}) provides a brief overview of the data collection routine and reduction process. Sec. \ref{sec:RESULTS} describes the asymmetries in the emission line profiles and the variabilities are discussed in Sec. \ref{sec:EW}. CI Tau's average 2022 profile is analyzed and compared to simulated models in Sec. \ref{sec:LINES}; first fitted to a circular disk model (Sec. \ref{sec:line_fitting}) and then a two-component disk model (Sec. \ref{sec:multicomponent}). Results from the profile fitting are compared to a hydrodynamic simulation in Sec. \ref{sec:hydro}. Sec. \ref{sec:diss} addresses lingering questions and how our results contribute to the debate around CI Tau's hot Jupiter companion. Lastly, Sec. \ref{sec:conclusion} highlights our main results.

\section{Observations and Reductions}
\label{sec:obs}
The spectra were collected with the iSHELL cross-dispersion echelle spectrograph at the NASA Infrared Telescope Facility (IRTF) \citep{2022PASP..134a5002R} for 11 of the 12 epochs in this study. Two of the epochs, from 03 Oct 2018 and 03 Jan 2019, were first included in \cite{2022AJ....163..174B} and the rest, obtained during the period of 21-29 Jan 2022, are presented in this work for the first time. In addition, archival data was downloaded for one epoch, 10 Oct 2008, from an older survey with NIRSPEC \citep{NIRSPEC} at the W. M. Keck Observatory. See Table \ref{tab:epochs} for an observation log. \par

In 2022, we observed CI Tau for 9 consecutive nights to cover one full orbital period of the companion CI Tau b. The spectra were acquired with an ABBA nodding pattern that allowed for the images to be combined in an A-B-B+A pattern. This removes sky emission to first order. The position angle of the slit on the sky was along the semimajor axis of the system's outer disk. For CI Tau we took 5 second exposures with 12 coadds while the telluric standard was observed with 5 second exposures with 5 coadds. For the iSHELL observations, the slit width of 0.375$\arcsec$ provided a spectral resolution of $\sim$92,000 \citep[][]{2022AJ....163..174B}. The average during the 2022 observation run was 0.84$\arcsec$. The NIRSPEC observations were acquired with a 0.432$\arcsec$ slit width that provided a spectral resolution of $\sim$25,000. \par

The iSHELL data were reduced using the SpeXtool5 pipeline \citep{2004PASP..116..362C, 2003PASP..115..389V} with which the calibration frames (sky frame and master flat) are prepared and the spectra orders are then straightened, extracted and wavelength calibrated using a sky model. The orders are then stitched together to construct the full spectrum. The telluric corrections are carried out by ratioing the wavelength calibrated science and standard stars. This cancels out the sky as well as stellar contributions. Utilizing a model generated by the Sky Synthesis Program \citep{Kunde1974}, the same process is followed to reduce the NIRSPEC data with a custom routine described in \cite{Brittain2003}. \par

\section{Results}
\label{sec:RESULTS}
\begin{figure*}[htb!]
    \centering
    \includegraphics[clip, trim = 0cm .5cm 15cm .5cm, width=0.85\textwidth]{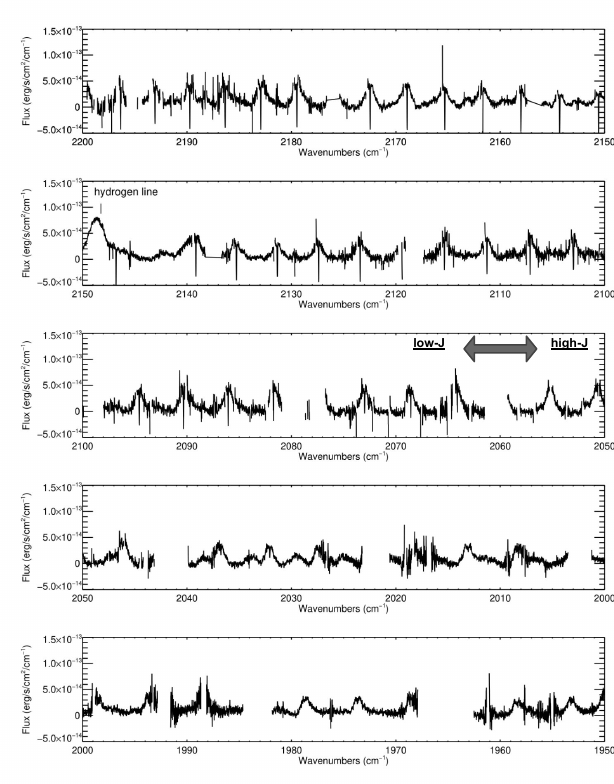}
    \caption{2022 average M$-$band spectrum of CI Tau (described in Sec. \ref{sec:RESULTS}). Captured are the ${}^{12}$CO emissions originating from the disk's surface. The low-J transitions ($\tilde{\nu}\geq$ 2058~cm$^{-1}$) have a superimposed narrow absorption feature indicative of lower temperature gas of the same species being present in the foreground. The most prominent feature is the hydrogen Pf$\beta$ line at 2147.8 cm$^{-1}$ which is used to infer stellar accretion rates (Sec \ref{sec:EW}).}
    \label{fig:full_Mband}
\end{figure*}

The average $M$-band spectrum from the 2022 observation run is presented in Fig. \ref{fig:full_Mband}. It covers the ${}^{12}$CO v=1-0 transitions from R(15)-P(42) as well as the hydrogen Pf$\beta$ transition. The low-J lines where J$\leq$20 (J being the rotational quantum number of the lower state) and $\tilde{\nu}_{\textrm{J}}\geq2060\rm~cm^{-1}$ (where $\tilde{\nu}_{\textrm{J}}$ is the central wavenumber of a transition) have a narrow absorption component superimposed on the broad emission features. This is an indication of foreground CO at a lower temperature than the emitting gas of the same species.

\begin{figure*}
    \centering
    \includegraphics[clip, trim = 0.5cm 0.45cm 5cm 0.5cm,width=0.90\textwidth]{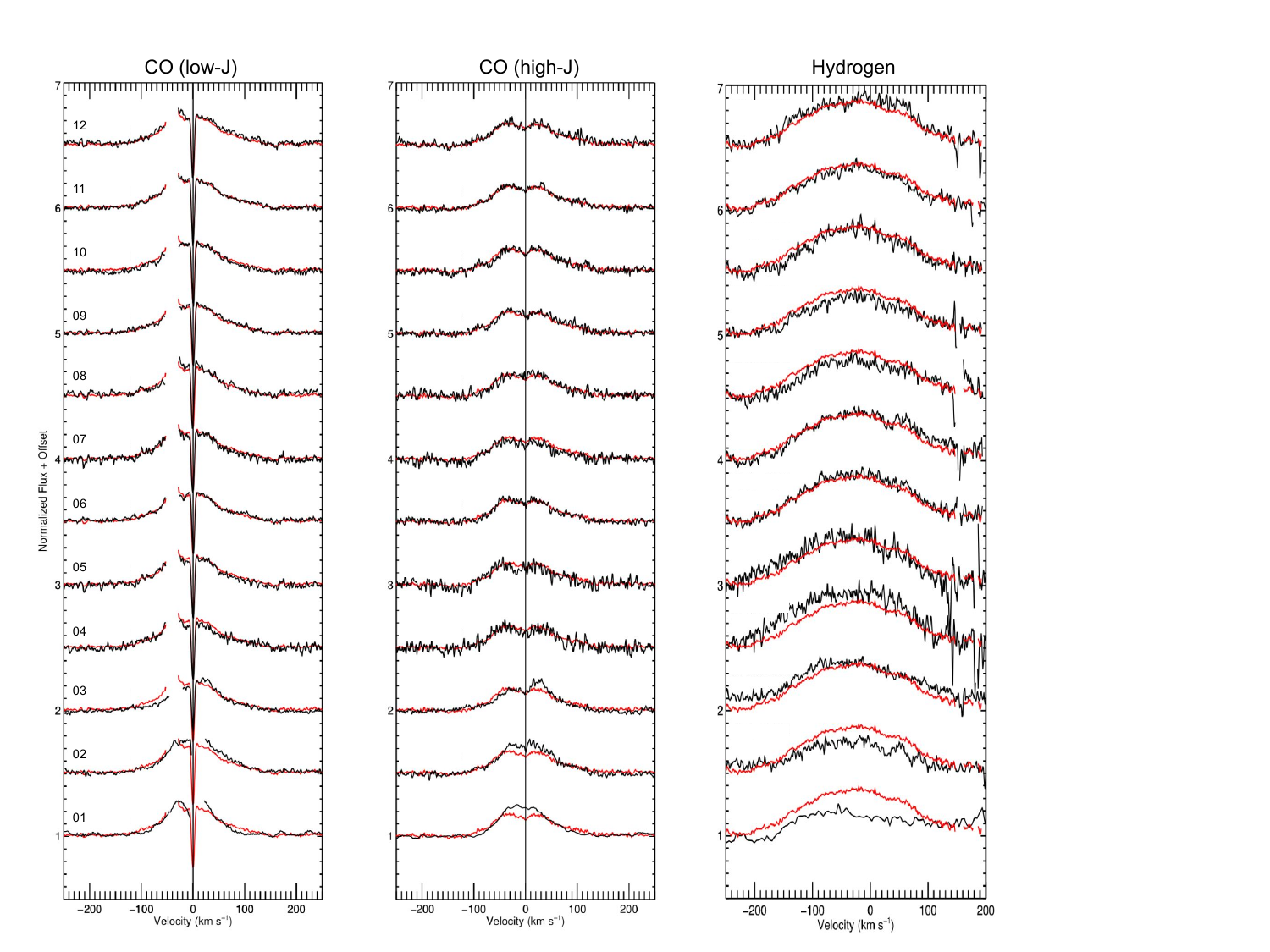}
    \caption{The normalized line profiles of CI Tau (described in Sec. \ref{sec:RESULTS}) with the 2022 average (red). The two left panels displays the ${}^{12}$CO low-J and high-J lines where The 2022 epochs (\#4-12) remained consistent but a clear variation is present when compared to the earlier epochs \#1-3. The right panel displays the hydrogen Pf$\beta$ line where variation was observed between every epoch.}
    \label{fig:all_profs}
\end{figure*}

The average emission line profiles of the ${}^{12}$CO low-J (J$\leq$20) lines, high-J (J$>$20) lines and the hydrogen Pf$\beta$ line are presented in Fig. \ref{fig:all_profs}. The epochs start at the bottom and progress upwards (the 2022 epochs begin with \#4) along with the 2022 average (red) plotted over each for comparison. To construct the average profile, each individual profile is adjusted so that their centers are aligned with each other in Doppler shifted velocity space (typically less then 0.5 km s$^{-1}$). This is done because the wavelength calibrations may not be the same between nights. The profiles are then shifted again to correct for their respective dates' Doppler motion to center the profiles at V = 0 km~s$^{-1}$. Finally, the average profile is calculated while excluding the profiles that have significant corruption. \par

The average stacked line profiles for the ${}^{12}$CO low-J lines are presented in the left column of Fig. \ref{fig:all_profs}. They possess an absorption component and a telluric contribution that we remove during the reduction pipeline-- this is seen as a gap in the profiles. Because of these external features the majority of the profile analysis is done on the high-J lines (center: Fig. \ref{fig:all_profs}).

Emission from a CPD would appear in the profiles as an asymmetric feature that oscillates over an orbital period with Keplerian velocities. The high-J line profiles (center; Fig. \ref{fig:all_profs}) displayed no clear asymmetric feature that varied over the 9~days. The lack of a variable feature places a limit on the emission from a CPD associated with CI Tau b and can be used to constrain the CPD radius.

To estimate the lower limit of our ability to detect CPDs, we estimate the minimum radius (R$^{\textrm{limit}}_{\textrm{CPD}}$) a CPD must have to be detected in our data. A best-fit temperature profile of the inner disk (Sec. \ref{sec:line_fitting}) indicates the temperature at CI Tau b's location is about 2300~K. This temperature is assumed to correspond to that of the CPD for simplicity. Using the M-band centered wavenumber (2127.7 cm$^{-1}$), the CPD's flux density is $B_{\rm J}(\tilde{\nu}_{\rm{J}},T) = 4.1\times10^{4}$ erg s$^{-1}$ cm$^{-2}$ ster$^{-1}$ cm, where $B_{\rm J}(\tilde{\nu}_{\rm{J}},T)$ is the Planck function.

The intensity of the CPD, $I_{\rm CPD}$, and $B_{\rm J}$($\tilde{\nu}_{\rm{J}}$,$T$) are related by $I_{\rm CPD} = \Omega B_{\rm J}(\tilde{\nu}_{\rm J},T) $, where $\Omega$ is the solid angle $\Omega = \pi (R_{\rm{CPD}}/d)^2$ and $d$ is the distance to the system (see Table \ref{tab:star}). We set $I_{\rm CPD}$ to our detection limit of 5$\sigma_{\textrm{F}}$,  where $\sigma_{\rm F} = 4.1 \times 10^{-15} ~\rm erg \,s^{-1} cm^{-2} cm$ is the standard deviation of the spectrum noise. Finally, $R^{\textrm{limit}}_{\textrm{CPD}}$ can be estimated as
\begin{equation}
\label{eq:Rlim}
    R^{\rm limit}_{\rm CPD} \approx d\sqrt{\frac{5\sigma_{F}}{\pi \rm B_{\rm J}(\tilde{\nu}_{\rm J},\textit{T})}} \,,
\end{equation}
(see also \cite{HD200546_CPD}). We find $R^{\rm limit}_{\rm CPD} =$ 0.013~au, or about 27 Jupiter radii. 

A CPD encompasses a fraction of its Hill radius ($R_{\rm H}$) and may be as large as $0.5R_{\rm H}$ \citep{Machida08}. Using the system parameters from Table \ref{tab:star}, and $a$ = 0.085~au, we obtain
\begin{equation}
    R_{\rm CPD} \approx 0.5 R_{\rm H} = 0.5 a(1-e) \left(\frac{M_{\rm P}}{3M_{*}}\right)^{\frac{1}{3}} \approx 0.005~{\rm au} \,.
\end{equation}
Since this value falls below our detection limit $R^{\rm limit}_{\rm CPD}$, we are not able to confirm whether the CPD around CI Tau b is present or not.

Although short term variability associated with a CPD was not observed in the 2022 epochs (\#4-12), there is a clear overall difference when compared to the earlier epochs (\#1-3). This is further discussed in Sec. \ref{sec:diss}.

The hydrogen Pf$\beta$ profiles are presented in the right column of Fig. \ref{fig:all_profs} with the 2022 average overplotted. Since this is an individual line at 2148.7 cm$^{-1}$, no averaging was performed. Because hydrogen probes regions much closer to the star their maximum velocities are much higher. These profiles also do not exhibit the double peaked structure that is often seen from disks. This transition suffers from a degree of blending because it occurs between the ${}^{12}$CO R(0) and R(1) lines. Because of this we limit the hydrogen profiles to velocities from V = -250 km s$^{-1}$ to +200 km s$^{-1}$. \par

Similar to the CO profiles, the earlier epochs of the hydrogen profiles differ from the 2022 ones. Unlike CO, though, they do display some variability through the consecutive 2022 epochs. Since hydrogen elucidates stellar behavior, this may be an indication of the companion modulating stellar accretion rates (see Sec. \ref{sec:EW}). \par

\subsection{Variability}
\label{sec:EW}
\begin{figure*}
    \centering
    \includegraphics[width=0.7\textwidth]{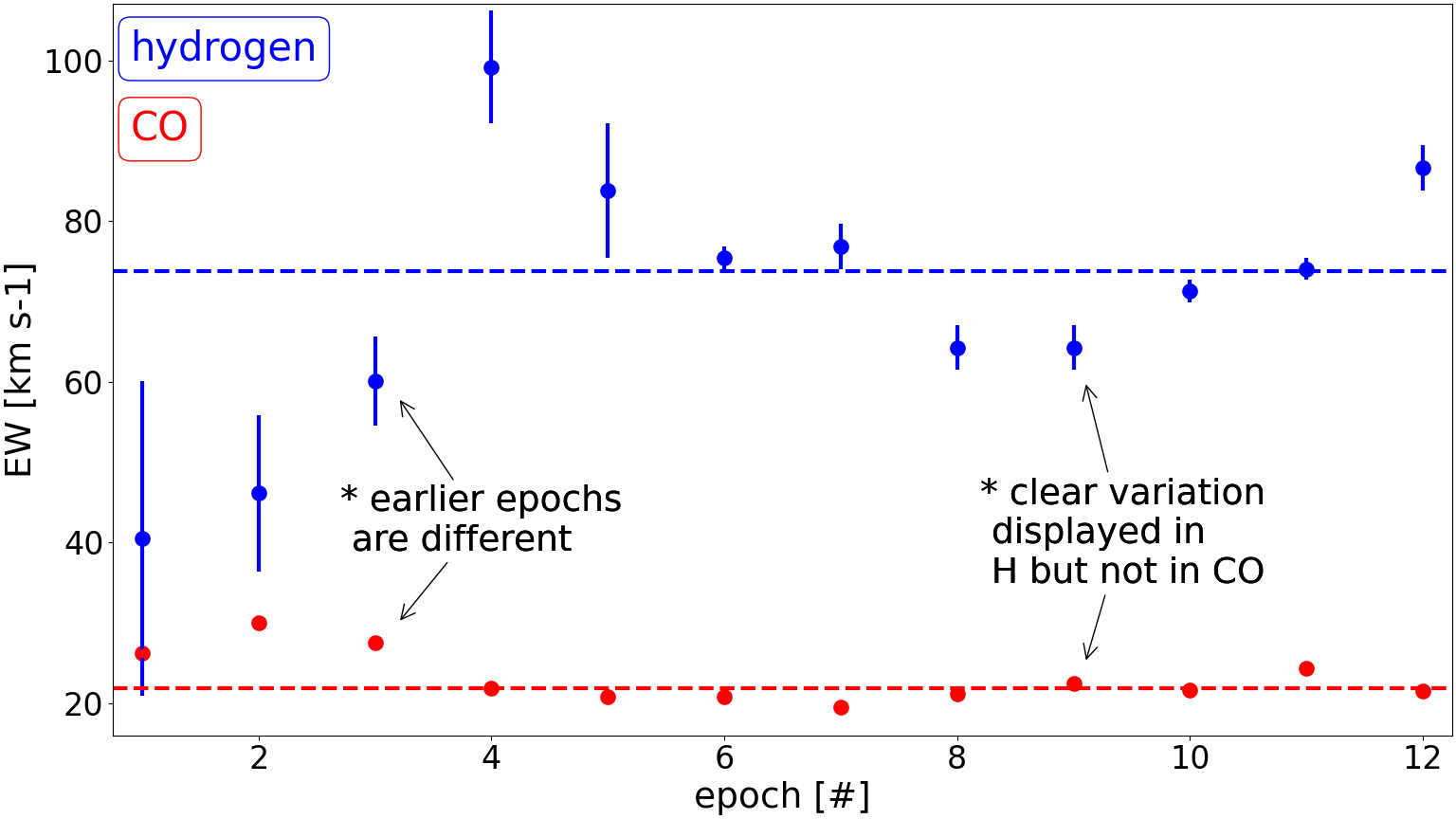}
    \caption{The equivalent widths (Sec. \ref{sec:EW}) of the average ${}^{12}$CO lines and the hydrogen line (see Table \ref{tab:co_eqw}) with the 2022 averages plotted as the horizontal dashed lines (${}^{12}$CO$_{\rm avg}$ = 21.8 km s$^{-1}$ \& H$_{\rm avg} = $73.7 km s$^{-1}$). In the 2022 epochs (\#4-12) we see the hydrogen vary with a possible 9~d period whereas the ${}^{12}$CO stayed consistent. The 2022 epochs are noticeably different from the earlier ones.}
    \label{fig:CO_EW}
\end{figure*}

The equivalent widths of the average line profiles of the CO transitions are presented in Table \ref{tab:co_eqw} and plotted as red points in Fig. \ref{fig:CO_EW}. The 2022 average value is 21.8 km s$^{-1}$ represented as a horizontal red dashed line. During the 2022 observation run, fluctuations about the average value are generally consistent with the measurement uncertainties of about 0.4 km s$^{-1}$; as a result, we have not detected any significant variability over the 9d period. We do note that epoch \#7 has a notably lower value of 19.46 km s$^{-1}$ while epoch \#11 has a higher value of 24.30 km s$^{-1}$. We consider them outliers but they may be physically meaningful if future observations find similar patterns. If there is a pattern over the course of an orbit then it is not clear. 

Overall, the 2022 measurements (\#4-12) are markedly lower than those seen in earlier epochs (\#1-3). This indicates changes in the disk over timescales much longer than the 9~d period of the 2022 run. This is discussed further in Sec. \ref{sec:diss}.

Also tabulated in Table \ref{tab:co_eqw} are the equivalent widths of the hydrogen Pf$\beta$ line alongside derived accretion rates (described later in this Sec.). The equivalent widths are plotted in Fig. \ref{fig:CO_EW} as blue points with the same units as CO. The hydrogen equivalent widths have a 2022 average of 73.7 km s$^{-1}$ and it is plotted as a horizontal dashed blue line. The equivalent widths do display a noticeable amount of variation over the 9~d period. Starting with the higher value at epoch \#4 with 99.2 km s$^{-1}$, the equivalent width then gradually dips to 64.3 km s$^{-1}$ by epochs \#8/\#9 and then rises again to 86.6 km s$^{-1}$ at epoch \#12. The small uncertainties in our measurements lend confidence to this trend being physical. 

The 2022 measurements are higher than the earlier measurements in general. Because the 2022 measurements are themselves variable and the earlier epochs have large uncertainties, it is inconclusive whether the prior epochs truly differ from the 2022 values. This is discussed further in Sec. \ref{sec:diss}.

Since we lack an absolute flux calibration, it remains uncertain during the 2022 observation run how much variability is attributable to the star itself. The stellar variability over the course of a day can be gauged by studying previous surveys. The AllWISE Multiepoch Photometry Database provides $4.6 ~\micron$ photometry with 21~hrs of coverage on 24 Feb 2010 and 24~hrs on 4 Sept 2010. The variation during those nights averaged 7\% and that is smaller than what is captured in the Pf$\beta$ line. Thus, fluctuations caused by the star are likely insufficient to explain the variability we observe in the Pf$\beta$ line. Nonetheless, a photometrically calibrated spectroscopic study would help clarify the nature of the observed variability.

The luminosity of the Pf$\beta$ line has been correlated to UV-excess-derived accretion luminosities and, thus, can be used as a reliable tracer of stellar accretion rates \citep{2013ApJ...769...21S}. The hydrogen equivalent widths are converted to their corresponding line luminosities utilizing the ALLWISE photometric data available and the GAIA DR3 distance. We do not include any extinction correction because it is expected to be negligible. While an M-band extinction coefficient, $A_{\rm M}$, for CI Tau is not available, one may estimate it using the J-band extinction coefficient, $A_{\rm J}$, of 0.51 reported by \cite{Kenyon&Hartmann1990}. Assuming ISM-like grain properties, we have $A_{\rm M}/A_{\rm J} = 0.095$ \citep{Mathis1990}, which translates to  $A_{\rm M} = 0.048$, or a flux correction of only about 4\%.

The Pf$\beta$ line luminosity ($L_{\textrm{Pf}\beta}$) is converted to an accretion luminosity ($L_{\textrm{acc}}$) using the relation given by \cite{2013ApJ...769...21S}:
\begin{equation}
    \log\left(\frac{L_{\rm acc}}{L_\sun}\right) = 0.91\log\left(\frac{L_{\rm Pf\beta}}{L_\sun}\right) + 3.29 \,,
\end{equation}
where we have elected to drop the uncertainties on the numerical constants as they are largely systematic and only serve to shift all data points equally. It is more meaningful to look at the uncertainties due to the measured equivalent widths alone.

The accretion luminosities are converted to stellar accretion rates ($\dot{M}$) by equating $L_{\rm acc}$ to the release of gravitational energy from $R_{\rm cor}$, the co-rotation radius in the disk, to $R_{*}$, the host star's radius:
\begin{equation}
    \dot{M} = \frac{R_{*}L_{\rm acc}}{\textrm{G}M_{*}} (1 - R_{*}/R_{\rm cor})^{-1} 
    \label{eq:star_acc} \, .
\end{equation}
The $(1 - R_{*}/R_{\rm cor})^{-1}$ reduces to a factor of 1.25 when $R_{\rm cor}$ is assumed to be $5R_{*}$ \citep{Hartmann1998}. We find an average accretion rate of $\dot{M}$ = 3.1$\pm$2.2 $\times$10$^{-8}$~$M_{\sun} ~\rm yr^{-1}$ which agrees with the previously reported value of $\dot{M}$ = 2.5$\pm$1.8 $\times$10$^{-8}$~$M_{\sun} ~\rm yr^{-1}$ \citep{2020MNRAS.491.5660D}.

The accretion rate varied by $35\%$ over the 9~d orbital period. Comparing to theoretical hydrodynamic models by \cite{2020MNRAS.495.3920T}, this level of variability can be caused by a companion with a mass of 9.4~M$_{\rm J}$ and an orbital eccentricity of about 0.05. Although that eccentricity is lower than that reported for CI Tau \citep{2019ApJ...878L..37F}, it is not beyond the realm of possibility given the large uncertainty in the reported eccentricity. Also, one should bear in mind that \cite{2020MNRAS.495.3920T} studied the disk accretion rate, which may have quantitative differences from the stellar accretion rate. While the companion may drive strong modulation in the flow of gas near its orbit, for that gas to travel further inward and ultimately accrete onto the star, it must still be subjected to some disk transport mechanisms. Transport by turbulence, for example, is diffusive and would likely smooth out the accretion flow. Detailed modeling \citep[e.g.,][]{ORBIT} is needed to verify whether the modulation we have detected here is consistent with the measured parameters of CI Tau b.

We estimate an upper limit to CI Tau b's accretion rate ($\dot{M}^{\rm limit}_{\rm p}$) by applying Eqn. \ref{eq:star_acc} where the $(1 - R_{*}/R_{\rm cor})^{-1}$ factor is now 1 by assuming, for simplicity, the gas falls onto the planet from a distance much larger than the planet's radius. We can write
\begin{equation}
    \dot{M}^{\rm limit}_{\rm p}= \frac{R_{\rm p}}{\textrm{G}M_{\rm p}}L^{\rm limit}_{\rm acc} \,,
    \label{eq:M_acc_p}
\end{equation}
where we choose $R_{\rm p}$ to be 2 Jupiter radii. According to \cite{Spiegel2012,Spiegel2013}, a 2 Myr giant companion with a mass of 10 M$_{\rm J}$ will have a radius of 1.94 R$_{\rm J}$ or 1.14 R$_{\rm J}$ depending on how it formed. Since CI Tau b has a slightly larger mass we used a value of 2 R$_{\rm J}$ for simplicity.

L$^{\rm limit}_{\rm acc}$ in Eqn. \ref{eq:M_acc_p} is related to our 5$\sigma_{\rm F}$ detection limit (see the sixth paragraph in Sec. \ref{sec:RESULTS}), or
\begin{equation}
    L^{\rm limit}_{\rm acc} \approx 5\sigma_{\rm F}  4\pi d^{2}  \Delta V \, ,
    \label{eq:L_acc_p}
\end{equation}
where $\Delta V$ is the width of the planet's hydrogen line. Assuming the planet's magnetospheric accretion flow speed is of order the free fall speed $\sqrt{GM_{\rm p}/R_{\rm p}} \approx 100 \rm~km~s^{-1}$ and the line width is roughly two times that, we get $\Delta V \approx 200 ~\rm km~s^{-1}$ or about 1.3 cm$^{-1}$ in wavenumber units. Finally, plugging these values into Eqn. \ref{eq:M_acc_p} and \ref{eq:L_acc_p}, the upper limit for CI Tau b's accretion rate is about $1\times10^{-11}$ $M_\sun$ yr$^{-1}$, or $1\times10^{-8}$ $M_{\rm J}$ yr$^{-1}$.

\subsection{Emission Line Analysis}
\label{sec:LINES}
\subsubsection{Circular Disk Model}
\label{sec:line_fitting}
We begin our analysis of the emission lines by assuming a circular disk. The spectrum (Fig. \ref{fig:full_Mband}) is fitted with a simple two-dimensional slab model that treats the emission as if it arises from a geometrically thin disk. A radial grid is defined from $r$=[$r_{\rm in}$,$r_{\rm out}$] which translates to the Keplerian velocity
\begin{equation}
    V_{\rm K}(r) = \sqrt{\frac{GM_{*}}{r}} \,.
\end{equation}
We choose to make the step sizes in the radial grid correspond with a step size of 1~km s$^{-1}$ in Keplerian velocities. This way the grid has a finer resolution closer to the star where the disk is more luminous. At each radius, the annulus is further divided by an angular array bounded from $\theta$=[0,2$\pi$]. $\theta$ is defined as the angle from the line-of-sight that has been projected onto the plane of the disk. Similarly, we made the angular step sizes correspond to a change in the projected velocities of 1~km s$^{-1}$. The projected velocities are
\begin{equation}
    V_{\rm p}(r,\theta) = V_{\rm K}(r) \sin\theta \sin i \,,
\end{equation}
where $i$ is the system's inclination. The projected velocities are then made the center of normalized Gaussian profiles which are represented as
\begin{equation}
    G(V-V_{\rm p}) = \frac{1}{b\sqrt{\pi}} \exp\left(\frac{-(V-V_P)^2}{b^2}\right) \,.
    \label{eq:gaussian}
\end{equation}
Here the Gaussian line width, $b$, is given the value $8.8$ km s$^{-1}$ to emulate a realistic level of blending between discrete projected velocities.

Each annulus corresponds to a radially dependent temperature and surface number density. Assuming basic power laws, these are expressed as
\begin{equation}
    T(r)=T_0\left(\frac{r}{r_{\rm in}}\right)^{\alpha} \,,
    \label{eq:temp}
\end{equation}
and
\begin{equation}
    N(r)=N_0\left(\frac{r}{r_{\rm in}}\right)^{\beta} \,,
    \label{eq:surden}
\end{equation}
Assuming thermodynamic equilibrium, the populations of the ro-vibrational states are
\begin{equation}
    N_{\rm J}= \frac{Ng_{\rm J}}{Q(T)} \exp\left(\frac{-E_{\rm J}}{k_{\rm B}\it T}\right) \,,
    \label{eq:NJ}
\end{equation}
where $E_{\rm J}$ and $g_{\rm J}$ are the energy and degeneracy of state J, respectively. $Q(T)$ is the partition function as a function of temperature.

The optical depths, $\tau_{\rm J}$, are
\begin{equation}
    \label{eq:tau}
    \tau_{\rm J}=\frac{N_{\rm J}}{8b\pi^{\frac{3}{2}}}\frac{g_{\rm J}A_{\rm J}}{g^{\prime}_{\rm J}\tilde{\nu}_{\rm J}^3} \,,
\end{equation}
where $A_{\rm J}$ and $\tilde{\nu}_{\rm J}$ are the Einstein $A$ coefficient and central wavenumber of the transition. $g_{\rm J}^{\prime}$ is the degeneracy of the adjacent state.

The flux densities are
\begin{equation}
    F_{\rm J} = (1-e^{-\tau_{\rm J}}) \,\frac{\tilde{\nu}_{\rm J}}{\rm c} \,\textrm{B}_{\rm J}(\tilde{\nu}_{\rm J},T) \,,
\end{equation}
where c is the speed of light in a vacuum, and B$_{\rm J}(\tilde{\nu}_{\rm J},T)$ is the Planck function expressed in units of wavenumbers ($\rm cm^{-1}$). The Planck function is evaluated at the central wavenumber of each transition, $\tilde{\nu}_{\rm J}$, and at the local temperature, $T$. The factor $\tilde{\nu}_{\rm J}/$c converts the units to Doppler-shifted velocities (km s$^{-1}$).

Finally, the line profiles, $I_{\rm J}(V)$, is obtained by convolving the emission from every grid point with $G(V-V_{\rm p})$ and integrating over the full disk, or
\begin{equation}
\label{eq:Iv}
    I_{\rm J}(V) = f_{\rm norm}\int_0^{2\pi} \int_{r_{\rm in}}^{r_{\rm out}} F_{\rm J}(r) \,\textrm{G}(V-V_{\rm p}(r,\theta)) \,r \,{\rm d}r \,{\rm d}\theta \,,
\end{equation}
where $f_{\rm norm}$ normalizes the $I_{\rm J}(V)$ to our observed line profiles. The profiles of the high-J lines are then averaged together and compared to CI Tau's profile (top left panel of Fig. \ref{fig:profile_flip}). We perform a $\chi^2$ optimization over six parameters: $T_0$ and $\alpha$ in the temperature profile (Eqn. \ref{eq:temp}); $N_0$ and $\beta$ in the surface density profile (Eqn. \ref{eq:surden}); $r_{\rm in}$; and $r_{\rm out}$. Our best-fit has a reduced $\chi^2$ value of 3.3.

The best-fit parameters are presented in Table \ref{tab:synpar}. We find that the temperature and surface density at the companion's location ($r=0.085$~au) are 2297 K and 2.9$\times10^{19}$ cm$^{-2}$, respectively. The radial extent of the disk goes from $r_{\rm in}$ = 0.048 $\pm$ 0.001~au to $r_{\rm out}$ = 0.898 $\pm$ 0.001~au.

In the top left panel of Fig. \ref{fig:profile_flip} we see that the circular disk model does broadly describe the observed line structures. However, when comparing just the red side of the profiles ($V > 0 \rm~km ~s^{-1}$), we see velocities where the model overshoots the data ($V \sim \rm +50~km ~s^{-1}$) but then undershoots it ($V \sim \rm +80~km ~s^{-1}$). This is peculiar because it cannot be matched even if we assign an eccentricity to the disk. Applying an eccentricity shifts the profile towards one direction effectively raising/lowering the entirety of one side to create a typical asymmetric line profile, but it cannot simultaneously increase and decrease the fluxes at different velocities on the same side. This discrepancy motivates us to propose a more complex disk model in the following section.

\subsubsection{Multi-component Disk Model}
\label{sec:multicomponent}
\begin{figure*}
    \centering
    \includegraphics[width=0.95\textwidth]{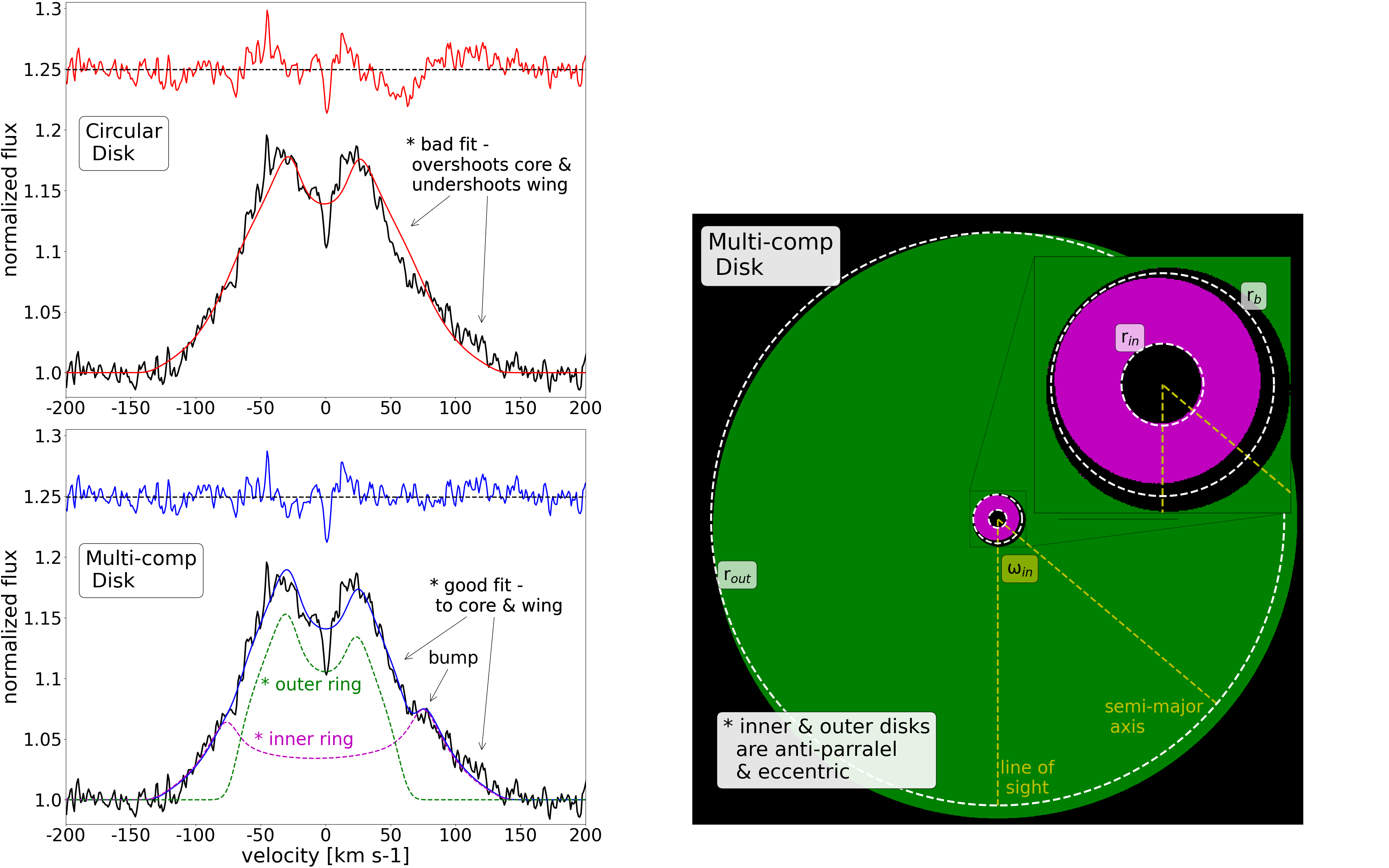}
    \caption{The 2022 average stacked emission line profile of the high-J ${}^{12}$CO lines is plotted with best-fit synthetic data alongside residuals (Sec. \ref{sec:LINES}). The top-left panel is a circular disk (Sec. \ref{sec:line_fitting}) fit that fails to replicate certain asymmetric features (see Table \ref{tab:synpar} for the best-fit parameters). The bottom-left panel is the multi-component disk fit (Sec. \ref{sec:multicomponent}) which allowed for the inner+outer components of the disk to contribute independently and provide a much better fit (see Table \ref{tab:synpar} for the best-fit parameters). The right figure is a visual of the best-fit multi-component disk where the inner+outer components are eccentric and anti-parallel.}
    \label{fig:profile_flip}
\end{figure*}

CI Tau's profile is better modeled when its disk is instead composed of multiple components each having their own eccentricities and arguments of periapses. This is motivated by the fact that the line core appears blue-shifted while the wings are red-shifted (Fig. \ref{fig:profile_flip}). If CI Tau's disk is divided into inner and outer components, the outer component can account for the blue-shifted core while the inner component accounts for the red-shifted wings of CI Tau's profile.

In this two-component model, we introduce a break radius, $r_{\rm b}$, that separates the inner and outer components. Within this radius the inner component will have an argument of periapse $\omega_{\rm in}$ that is different from that of the outer component $\omega_{\rm out}$. In our model, $\omega_{\rm in}$ is defined to be the angle between the inner disk's semi-major axis and our line-of-sight projected onto the plane of the disk. The disks are required to be oppositely oriented to some degree in order to replicate CI Tau's profile (Fig.\ref{fig:profile_flip}) so, to reduce the number of free parameters, we chose to fix them to be anti-parallel ($\omega_{out} = \omega_{in} + \pi$). The validity of this assumption is discussed in Sec. \ref{sec:hydro}.

The two components are also given their own eccentricities, $e_{\rm in}$ and $e_{\rm out}$, which are treated as constants throughout their respective component. We assign a width parameter $\Delta$ around the break radius where neither disk will be contributing flux. This may capture any potential disk gap that might be present.

Unlike in Sec. \ref{sec:line_fitting} where we fitted the overall spectrum with a temperature and surface density profile (Eqns. \ref{eq:temp} and \ref{eq:surden}) as a means to compute the disk's intensity (Eqn. \ref{eq:Iv}), here we directly fit the intensity as a function of radius $I(r)$. Since we are now fitting only to the high-J lines there is not enough information to be able to constrain the temperature and surface density. $I(r)$ is fitted as a broken-power law that follows one exponent for the inner component $\iota_{\rm in}$ and another one for the outer component $\iota_{\rm out}$. The broken power law is defined as follows:
\begin{equation}
    \label{eq:broke}
    I(r) = I_0 \left(\frac{r}{r_{\rm b}}\right)^{-\iota_{\rm in}} \left(\frac{1}{2}\left(1+\left(\frac{r}{r_{\rm b}}\right)^{\frac{1}{\delta}}\right)\right)^{(\iota_{\rm in}-\iota_{\rm out})\delta} \,,
\end{equation}
where $I_0$ is a normalization constant. We fix $\iota_{\rm in}=0$ because the inner disk is likely narrow and it helps simplify our fit. $\delta$ determines the ``smoothness'' of transition between power-laws. We set it to be $\delta = \log(1+\Delta/r_{\rm b})$ in order to emulate a smooth transition that occurs over the distance $\Delta$. In the intensity profile, $\iota_{\rm out}$ is the only parameter that we vary.

Since our model now considers eccentricity, it is simpler to operate in semi-major axis space, where the distance to the star is now dependent on the semi-major axis, eccentricity and azimuthal angle:
\begin{equation}
    r = \frac{a(1-e^2)}{1+e\cos(\theta-\omega)}\,,
\end{equation}
where the variables $e$ and $\omega$ are $e_{in}/e_{out}$ or $\omega_{in}/\omega_{out}$ for the inner/outer components, respectively. Like with the circular disk model, $\theta$ is the angle between the grid element and the line-of-sight in the disk's plane.

The projected velocities also change depending on the eccentricity and argument of periapse of the annulus. The new expression for the projected velocities is
\begin{equation}
    V_{\rm P}(a,\theta) = \sqrt{\frac{\textrm{G}M_{*}}{a(1-e^2)}} (\sin\theta+e\sin\omega)\sin i \, .
\end{equation}

An $I(r,\theta)$ is calculated for every grid element and transformed into a line profile in velocity space by multiplying it with the Doppler-shifted normalized Gaussian profile described in Eqn. \ref{eq:gaussian}. This is done over the extent of the disk, summed together, and normalized to CI Tau's profile by adjusting $I_0$ in Eqn. \ref{eq:broke}.

Our $\chi^2$ optimization finds a best-fit that has a reduced $\chi^2$ of 0.984. This is a significant improvement over the circular disk model that has a reduced $\chi^2$ of 3.3 (Sec. \ref{sec:line_fitting}). The best-fit profile of the multi-component disk is displayed in the bottom left panel of Fig. \ref{fig:profile_flip} together with the contributions from the inner and outer disks. 

We confirm that the multi-component model can replicate the blue-shifted line core and the red-shifted wings seen in CI Tau's profile; which is not possible for a circular disk. The ``bump'' at V = 80 km s$^{-1}$ is an exaggerated by-product of our model because of the lack of flux from the region centered on the discontinuity. A more detailed analysis can be performed in the future where this is filled with some realistic emission that should aid in smoothing out the 'bump' and provide an even better fit.
 
The best-fit parameters for the multi-component disk are presented in Table \ref{tab:synpar}. The CO extends from $r_{\rm in}$ = 0.052$\pm$0.002~au to $r_{\rm out}$ = 1.53$\pm$0.16~au. Compared to the circular disk (Sec. \ref{sec:line_fitting}), the inner radius marginally changed while the outer radius increased by 70\%. The inner radius is also close to the truncation radius estimated by \cite{Gravity23} ($0.034\pm0.014 \rm~au$) and may extend to within the star's co-rotation radius
\begin{equation}
\label{rcor}
    R_{\rm rot} = \left(\frac{\textrm{G}M_{*} P_{\rm rot}^2}{4 \pi^2}\right)^{\frac{1}{3}} \,.
\end{equation}

Assuming Keplerian rotation, we calculate the co-rotation radius to be 0.069~au. Similar results have been found for other T Tauri stars \citep[e.g.,][]{Carr_probe}. The same calculation for CI Tau's co-rotation radius was done by \cite{Gravity23} but there they assumed a different rotational period and stellar mass.

The best-fit value for the eccentricities of the inner and outer components are $e_{\rm in}$ = 0.056 $\pm$ 0.015 and $e_{\rm out}$ = 0.048 $\pm$ 0.008. These values can be seen as luminosity-weighted averages that are more representative of the inner edges of the components, which are more luminous. Overall, these eccentricities are much smaller than that of CI Tau b's orbit ($e$ = 0.25; \cite{2019ApJ...878L..37F}). This may be an indication of the planet's orbit being less eccentric, since one might expect the planet's and disk's eccentricities to be similar. This is discussed further in Sec. \ref{sec:hydro}.

Our best-fit gives $\Delta = 0.009 \pm 0.003$ au, which is the half-width of the disk break. As a reminder, it also represents a region in the disk (i.e., a ``gap'') where we assigned no emission. This is about 6\% of $r_{\rm b}$ = 0.14~au, the location of the break, which implies that the transition occurs sharply. Taking into account the eccentricities of the disks, the full width of the gap represented by $\Delta$ varies between 0.0035 au at the apoapsis (periapsis) of the inner (outer) disk, to 0.032 au at the periapsis (apoapsis) of the inner (outer) disk. Considering the uncertainties in our fit, this gap is too narrow for us to definitively confirm its existence. It mainly helps to separate the two disks just enough to avoid overlapping (see the right panel of Fig. \ref{fig:profile_flip}).

We find the argument of periapsis of the inner disk ($\omega_{\rm in}$) to have an upper limit at 40$\degr$. Again this describes the angle between the semi-major axis of the inner component and the line-of-sight projected onto the plane of the disk. This would correspond to the outer disk having a periapsis of $\omega_{\rm out}=\omega_{\rm in}+\pi = 220\degr$. 

The outer component's intensity profile follows a $\iota_{\rm out} =$ 3.11 dependence. These values are positive because a negative sign is incorporated in Eqn. \ref{eq:broke}. The large value of $\iota_{\rm out}$ indicates a steep drop in flux---a sign that a large portion of the disk's emission originates from the inner component. 

We would like to remind the reader that our results are sensitive to the mass of the star and inclination of the disk. Here we assume an inclination of $i$ = 71\degr \citep{Gravity23} and a stellar mass of $M_{*} = 1.02 ~M_{\sun}$ \citep{Law2022} but these parameters have changed multiple times now. For example, the mass of the star was originally 0.8 $M_{\sun}$ \citep{Guilloteau2014} but then it was updated once to 0.9 $M_{\sun}$ \citep{Simon2019} and then again to the current value we used. It is possible that these values can be updated further and, if they do, then our best-fit radii would change by the following factor
\begin{equation}
    \label{eq:rnew}
    \frac{r_{\rm new}}{r_{71\degr}} = \left(\frac{\sin(i_{\rm new})}{\sin(71\degr)}\right)^{2} \left(\frac{M_{\rm new}}{1.02 M_{\sun}}\right) \, ,
\end{equation}
where $r_{71\degr}$ corresponds to the values assigned to the $r_{\rm in}$, $r_{\rm out}$ and $r_{\rm b}$ parameters. Changing these quantities by a constant factor will have no qualitative effect on our overall disk model.

\subsection{Hydrodynamic Simulation}
\label{sec:hydro}
In this section, we present a proof-of-concept simulation that qualitatively demonstrates how a planet might be able to generate the disk features inferred by our model. The simulation is performed using PEnGUIn \citep{Fung2015} with a setup similar to the one used by \cite{PDS70}, but here the planet's mass ($M_\textrm{P}$) and orbit ($a$ and $e$) are fixed. We pick a representative planet mass $M_\textrm{P}$ = 1~M$_{\rm J}$, and assign it an orbital semi-major axis that equals $r_{\rm b}$ in our multi-component model (Table \ref{tab:synpar}), and an orbital eccentricity of $0.05$, similar to the eccentricities of both the inner and outer disk. We choose not to use a planet mass as large as the proposed CI Tau b's mass (11.6 $M_{\rm J}$) because \cite{Kley2006} had previously demonstrated that planets more massive than about 5 $M_{\rm J}$ would excite disk eccentricities much larger than those inferred by our model.

\begin{figure*}
    \centering
    \includegraphics[width=0.85\textwidth]{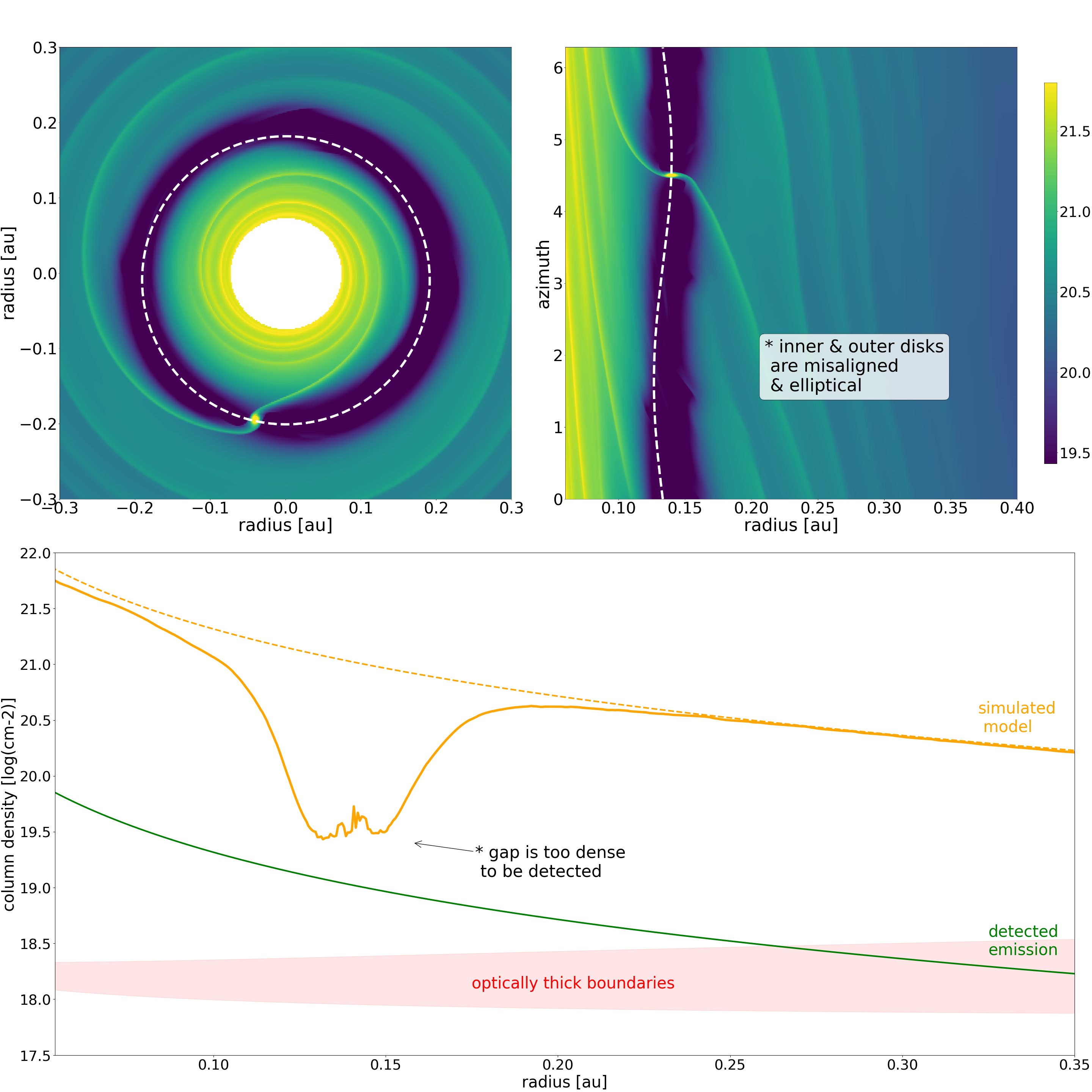}
    \caption{The top row is the final snapshot of the hydrodynamic simulation (Sec. \ref{sec:hydro}) with a hypothetical $1 M_{\rm J}$ planet that may reproduce the disk eccentricities we observed (0.05; Fig. \ref{fig:profile_flip}). The color scale is logarithmic and in units of CO molecules cm$^{-2}$. The left and right panels are the same picture in Cartesian and polar coordinates, respectively. The white dashed line traces the orbit of the simulated planet. In this case, the outer disk develops an eccentricity similar to the planet's, while the eccentricity of the inner disk is more subtle. The bottom row plots the azimuthally averaged CO column density profile. The orange solid curve is taken from the same simulation and time-averaged over the last orbit; the orange dashed curve represents the unperturbed disk; and the green solid curve is the emission we detected (Table \ref{tab:synpar}; Sec. \ref{sec:line_fitting}). A $\tau = $1 region was calculated for the P(20)-P(40) transitions and plotted as the red region (Sec. \ref{sec:hydro}). The simulated profile (orange) is normalized so that, even at the bottom of the planetary gap, it lies above our detected CO column, illustrating that even though the gap is depleted by a factor of $\sim$100, its ${}^{12}$CO emission can remain optically thick (Sec. \ref{sec:hydro}).}
    \label{multicomp}
\end{figure*}

The simulation is locally isothermal and follows Eqn. \ref{eq:temp} and Table \ref{tab:synpar} for the temperature profile. The surface density has an initial power-law profile the same as Eqn. \ref{eq:surden} with $\beta=-2$, but the normalization is set to 1 in code units. Because we do not consider the self-gravity of the disk, the normalization to the surface density profile plays no role in the dynamics simulated. In Fig. \ref{multicomp}, we assign a physical value to the normalization based on the fact that we did not detect a planetary gap (Sec. \ref{sec:multicomponent}). Also we assume a constant CO/H$_{2}$ ratio. The Sunyaev-Shakura viscosity parameter is assumed to be $0.01$. Our choice describes a turbulent disk, which is plausible at short distances from the star where the ionization fraction is expected to be high and the magnetorotational instability (MRI; \citealt{MRI1991}) is expected to be active; the choice of $0.01$ is roughly similar to the most turbulent MRI-active disks \cite[e.g.,][]{Simon2009,Guan2009}.

The final snapshot of the hydrodynamic simulation after 1000 planetary orbits is presented in Fig. \ref{multicomp}. The top left and top right panels are in Cartesian and polar coordinate systems, respectively. The inner and outer disks have developed a subtle eccentricity that, generally, appears similar in magnitude to the planet's eccentricity, but only near the planet's orbit. A slightly higher planet mass and/or a higher planetary eccentricity might better match our observation. However, the disk would develop an eccentricity that is too high if the planet was given the same parameters as CI Tau b.

Although the arguments of periapses of the inner and outer disks do not appear to be exactly anti-parallel as they were in our model (Fig. \ref{fig:profile_flip}), they are far from aligned. The precise geometry of the inner and outer disks is complicated. While parts of the disk further away from the planet's orbit may be approximated by eccentric orbits \citep{Kley2006}, gas streamlines that co-orbit with the planet should follow horseshoe orbits that are deformed by the planet's eccentricity \citep{Pan04}. Even though our slab model does not capture all the nuances, we are confident that it does capture the fact that the inner and outer disks have distinct geometries.

In this simulation, the planet carves out a gap that is depleted in gas by about two orders order of magnitude and about $0.05$ au in width. If the disk is sufficiently dense, the CO emitting layer can remain undepleted, producing the optically thick emission we observed, even inside the gap; but what defines ``sufficiently'' dense is a complicated matter that involves not only the H/CO ratio, but also the expected abundances of CO at low column densities \citep[e.g.,][]{Bruderer2013,Doppmann2017}. To estimate how much CO is needed to produce optically thick emissions, we first set $\tau_{\rm J}=1$ in Eqn. \ref{eq:tau}. From there, we obtain the ro-vibrational populations $N_{\rm J}$, which is then converted to the CO surface density, $N$, using Eqn. \ref{eq:NJ}. In the bottom row of Fig. \ref{multicomp} we plot the $\tau_{\rm J}=1$ region (``optically thick boundaries'') for the high-J P(20)-P(40) lines. The azimuthally averaged column density profile of the simulation (orange: Fig. \ref{multicomp}) is scaled so that it lies just above the $\tau_{\rm J}=1$ regime and the detected CO surface density profile (green). This represents a possible version of the disk where a planet has carved out a gap that our observation is not sensitive to, though it is possible for the disk to have an even higher density.

Taking this further, we can see what constraint this assumption implies for the disk mass. In the bottom of Fig. \ref{multicomp}, we see that the unperturbed profile (orange dashed) of the simulated model needs to be about two orders of magnitude higher than Eqn. \ref{eq:surden}, the column density of the emitting layer. Or, more precisely, the full column density may be expressed as $\Sigma = 9.0\times10^{21} (r/r_{\rm in})^{-2} \rm~cm^{-2}$. The total number of CO molecules can then be calculated by integrating this over the extent of the disk. Or,
\begin{equation}
    N_{\rm CO} = 2\pi\int_{r_{\textrm{in}}}^{r_{\textrm{out}}} 9.0\times10^{21}~{\rm cm^{-2}} \left(\frac{r}{r_{\rm in}}\right)^{-2} r \,{\rm d}r \, .
\end{equation}
Using the parameters from Table \ref{tab:synpar}, we get $N_{\rm CO} \sim 10^{47}$ CO molecules. Assuming a CO/H$_2$ ratio of 1.6$\times10^{-4}$ \citep{France2014COH2AR}, this translates to a mass of about $10^{-6}$ $M_{\sun}$ or $10^{-3}$ $M_{\rm J}$. This low disk mass may help explain why the planet can remain on an eccentric orbit over many orbital periods---eccentricity damping due to planet-disk interaction scales linearly with disk gas density \citep{artymowicz92}. Future detailed modeling can better evaluate the strength of eccentricity damping and explore the implications on the planet's orbital evolution. Because this disk mass estimate uses a density profile that lies just above our detected CO column, it can be seen as a lower limit.

\section{Discussion}
\label{sec:diss}
The hot Jupiter companion CI Tau b was discovered by \cite{Johns2016} via IR radial velocity monitoring and \cite{2019ApJ...878L..37F} detected CO directly from the planet's atmosphere. These studies cross-validated the planet's orbital period to be 9~d. The host star's rotational period was found by \cite{biddle2018} to be 6.6~d through analysis of the K2 lightcurves. However, the origin of these signatures was questioned by \cite{2020MNRAS.491.5660D} who claimed that both signatures can be attributed to the star itself. A follow-up analysis of the K2 photometry by \cite{biddle2021} affirms that signature strengths cannot be replicated by stellar activity alone. Whether CI Tau b as a planet exists remains an on-going investigation.

One piece of our results that may validate CI Tau b is the seemingly 9~d period captured in the stellar accretion rate, derived from the hydrogen equivalent widths (Sec. \ref{sec:EW}). We also find some indirect evidence for a planet's existence (but not necessarily CI Tau b) through planet-disk interaction, where we find an inner and outer disk that have separate eccentricities and arguments of periapses (Sec. \ref{sec:multicomponent}). Further investigations are needed to explain some other features we observed. Below we list a few that we find the most puzzling.

As illustrated in Fig. \ref{fig:CO_EW}, we observed the hydrogen Pf$\beta$ equivalent widths to vary by 35\% during the 2022 epochs (\#4-12; Sec. \ref{sec:EW}). This variation also encompasses the values observed in epoch \#3 but not \#1-2. Since it is ambiguous how representative our data is of the overall behavior, it remains unclear if epochs \#1-2 are atypical or not. For instance, they are around 45\% from the 2022 mean and have relatively large uncertainties; statistically, it is possible that they are consistent. However, if they are meaningfully different then that may indicate changes in the star's luminosity and accretion rates. CTTSs are known to be quite variable over a wide array of time periods so CI Tau could have simply changed its luminosity or accretion rate between observation runs. Another possible explanation is that if the accretion flow is not isotropic, its direction might be related to the disk's eccentricity. Since eccentric disks driven by planet-disk interaction are known to precess over hundreds of planetary orbits \citep[e.g.,][]{Kley2006}, the direction of the accretion flow may change with it and we would observe a variable accretion luminosity as the accretion flow goes in and out of alignment with our light-of-sight.

Unlike hydrogen, the ${}^{12}$CO equivalent widths (Fig. \ref{fig:CO_EW}) appear less variable (Sec. \ref{sec:EW}). In 2022 they varied by 12\% and that, along with having much smaller uncertainties, leads us to believe that the earlier epochs (\#1-3), which are $>$20\% from the 2022 mean, are different. Similar to the discussion in the previous paragraph, this may be due to the disk precessing and altering the amount of emission directed along our line-of-sight. Changes in the disk's eccentricity was observed over many planetary orbits (Sec. \ref{sec:RESULTS}). However, there remains the possibility, again, that the star changed its luminosity and, as such, the temperature of the disk. If this were the case, then it means that from 2008 to 2022, the disk's luminosity decreased while the star's increased.

CI Tau's average emission line profile of 2022 was fitted with a disk that is constructed from two components (Fig. \ref{fig:profile_flip}) that have their own eccentricities (Sec \ref{sec:multicomponent}). Our derived eccentricities, which are about $0.05$ (Table \ref{tab:synpar}), are lower than that of CI Tau b, which is $e=0.25\pm0.16$. We also find that the disk has a very narrow gap if it exists; in other words, we find no evidence of gas at any radius traveling with an eccentricity as large as CI Tau b's. It is not a stable configuration for a planet to have a highly eccentric orbit and be simultaneously embedded in a near-circular disk---either the disk will damp the planet's eccentricity \citep[e.g.,][]{Duffell15} or the planet will force the disk to become more eccentric \citep[e.g.,][]{Bitsch13}. Given the large uncertainty in CI Tau b's eccentricity, it may be possible that its true value is only about $0.05$. According to the analysis by \cite{2020MNRAS.491.5660D}, it is also possible that CI Tau b does not exist, and here we are instead observing the influence of another planet with a much lower eccentricity, such as the one simulated in Sec. \ref{sec:hydro}.

\section{Summary and Conclusions}
\label{sec:conclusion}
CI Tau was observed for 9 consecutive nights with the NASA IRTF in January of 2022 (Sec. \ref{sec:obs}). This data was reduced (Fig. \ref{fig:full_Mband}) and paired with older data taken in 2008 using Keck and 2018/2019 also with IRTF, thus giving us a total of 12 epochs (Table \ref{tab:epochs}). For each epoch, we constructed the emission line profiles of the ${}^{12}$CO ro-vibrational transitions and the hydrogen Pf$\beta$ line (Fig. \ref{fig:all_profs}). We fitted CI Tau's average high-J line profile from 2022 with flat, two-dimensional disk models, where we considered both a circular model (Sec. \ref{sec:line_fitting}) and an eccentric model with two components (Sec. \ref{sec:multicomponent}). From these fits, we constrained the disk properties (Table \ref{tab:synpar}), such as the temperature (Eqn. \ref{eq:temp}) and surface density (Eqn. \ref{eq:surden}) profiles, and the eccentricities ($e_{\rm in}$ and $e_{\rm out}$) of the different components. Our main conclusions are highlighted below:
\begin{itemize}
    \item The core of the average stacked ${}^{12}$CO line profile is blue-shifted while the wings are red-shifted (Sec. \ref{sec:multicomponent}). This is well modeled by introducing an inner and outer disk component that are both eccentric but oppositely oriented (Fig. \ref{fig:profile_flip}). In our best-fit model, the components are separated near $r_{\rm b}$ = 0.14~au (Table \ref{tab:synpar}). Both components also have their own eccentricities of about $0.05$. The disk's structure may be explained by an embedded giant companion around $r_{\rm b}$ with a similar eccentricity (Sec. \ref{sec:hydro}).

    \item A 9~d variability was observed in the hydrogen Pf$\beta$ line (Fig. \ref{fig:CO_EW}), which indicates variations in the stellar accretion rate (Sec. \ref{sec:EW}). A 9~d periodicity has been reported before by multiple groups, who attributed it to the presence of a giant companion \citep{Johns2016,biddle2021} or stellar activity \citep{2020MNRAS.491.5660D}.

    \item The ${}^{12}$CO lines displayed varying asymmetries over hundreds of orbits (Fig. \ref{fig:all_profs}). This might indicate that the inner disk around CI Tau is eccentric and precessing (Sec. \ref{sec:RESULTS}).

    \item We did not detect ${}^{12}$CO emission from the circumplanetary disk around the proposed planet CI Tau b \citep{Johns2016,2019ApJ...878L..37F,biddle2021}, and showed that we were likely not sensitive to it (Sec. \ref{sec:RESULTS}). 
    We also did not detect hydrogen Pf$\beta$ emission from the planet, which allowed us to place an upper limit on the planet's accretion rate of 1$\times$10$^{-8}$ Jupiter-mass per year (Sec. \ref{sec:EW}).

    \item The inner radius of the protoplanetary disk was fitted to be $0.052\pm0.002$~au which is consistent with the truncation radius of $0.034\pm0.014$~au reported by \cite{Gravity23} and may extend within the star's corotation radius of 0.069~au.

    \item The average accretion rate of CI Tau in 2022 is 3.1$\times$10$^{-8}$ $M_{\sun}$ yr$^{-1}$. This aligns with what was previously reported by \cite{2020MNRAS.491.5660D} (2.5$\times$10$^{-8}$).

\end{itemize}

Being potentially the youngest host of a hot Jupiter companion, the CI Tau system can serve as the testing ground for theories pertaining to planet-disk interactions. Below, we chart a few directions to further investigate this system.

The CO and hydrogen equivalent widths (Table \ref{tab:co_eqw}; Fig. \ref{fig:CO_EW}) were observed to be markedly different in prior years. Future observations of the CI Tau system separated by a similar time span may be able to confirm and characterize this long-term behavior. Explaining it, such as whether it is caused by disk precession, would require detailed modeling.

Our analysis of the ${}^{12}$CO emissions (Sec. \ref{sec:LINES}) utilized a simple two-dimensional slab model that does not adequately account for three-dimensional effects such as disk flaring and vertical temperature variations. Future analysis may incorporate a more rigorous disk geometry that may capture more complexities.

The hydrodynamic simulation in this study (Sec. \ref{sec:hydro}) was a proof-of-concept for how an embedded giant planet may create the disk features observed. A more rigorous numerical study may better constrain the parameters of the planet-disk system.

\section{Acknowledgements}
We thank Jean-Francois Donati for a helpful discussion and providing input on the CI Tau system.
This work includes data gathered at the Infrared Telescope Facility, which is operated by the University of Hawaii under contract 80HQTR19D0030 with the National Aeronautics and Space
Administration.
Some of the data presented herein were obtained at the W. M. Keck Observatory, which is operated as a scientific partnership among the California Institute of Technology, the University of California and the National Aeronautics and Space Administration. The Observatory was made possible by the generous financial support of the W. M. Keck Foundation.
The authors wish to recognize and acknowledge the very significant cultural role and reverence that the summit of Maunakea has always had within the indigenous Hawaiian community.  We are most fortunate to have the opportunity to conduct observations from this mountain. 
\bibliography{sample631.bib}{}
\bibliographystyle{aasjournal}


\begin{deluxetable}{ccc}
\tablecaption{CI Tau system parameters \label{tab:star}}
\tablewidth{0pt}
\tablehead{
\colhead{parameter} & \colhead{value} & \colhead{ref.}}
\startdata
$d$           & 160 $\pm$ 10\rm~pc          & \cite{Gaia} \\
$P_{\rm orb}$ & 9.00 $\pm$ 0.5~d            & \cite{biddle2021}\\
$e$           & 0.25 $\pm$ 0.16             & \cite{2019ApJ...878L..37F} \\
$M_{\rm P}$   & 11.6 $\pm$ 2.8 $M_{\rm J}$  & \cite{2019ApJ...878L..37F} \\
$M_{*}$       & 1.02 $\pm$ 0.001 $M_{\sun}$ & \cite{Law2022} \\
$R_{*}$       & 2.0 $\pm$ 0.3 R$_{\sun}$    & \cite{2020MNRAS.491.5660D} \\
$i$           & 71 $\pm$ 1 $\degr$          & \cite{Gravity23} \\
$P_{\rm rot}$ & 6.62 $\pm$ 0.2~d            & \cite{biddle2021} \\
\enddata
\end{deluxetable}

\begin{deluxetable}{cccccc}
\tablecaption{Observation logs \label{tab:epochs}}
\tablewidth{0pt}
\tablehead{
\colhead{Epoch} & \colhead{JD} & \colhead{UT} & \colhead{Int. Time} & \colhead{Telluric} & \colhead{Coverage} \\
\colhead{\#} & \colhead{\_} & \colhead{\_} & \colhead{min} & \colhead{Standard} & \colhead{cm$^{-1}$}}
\startdata
1    & 2454743    & 10 Oct 2008    & 8   & HR 1251 & 1970-2150 \\
2    & 2458402    & 03 Oct 2018    & 114 & HR 1165 & 1950-2200 \\
3    & 2458487    & 03 Jan 2019    & 96  & HR 1791 & 1950-2200 \\
4-12 & 2459601-09 & 21-29 Jan 2022 & 90  & HR 1791 & 1950-2200 \\
\enddata
\end{deluxetable}

\begin{deluxetable}{ccccc}
\tablecaption{Equivalent widths and accretion rates \label{tab:co_eqw}}
\tablewidth{0pt}
\tablehead{
\colhead{Epoch} & \colhead{EW (CO)} & \colhead{EW (Pf$\beta$)} & \colhead{log($\dot{M}$) } \\
\colhead{\#} & \colhead{km s$^{-1}$} & \colhead{cm$^{-1}$} & \colhead{$M_{\sun} \rm~yr^{-1}$} }
\startdata
1  & 26.22$\pm$0.22 & 0.29$\pm$0.14 & -7.76$\pm$0.19 \\
2  & 29.96$\pm$0.30 & 0.33$\pm$0.07 & -7.71$\pm$0.08 \\
3  & 27.56$\pm$0.19 & 0.43$\pm$0.04 & -7.61$\pm$0.04 \\
4  & 21.80$\pm$0.75 & 0.71$\pm$0.05 & -7.41$\pm$0.03 \\
5  & 20.75$\pm$0.61 & 0.60$\pm$0.06 & -7.47$\pm$0.04 \\
6  & 20.78$\pm$0.27 & 0.54$\pm$0.01 & -7.52$\pm$0.01 \\
7  & 19.46$\pm$0.42 & 0.55$\pm$0.02 & -7.51$\pm$0.01 \\
8  & 21.11$\pm$0.56 & 0.46$\pm$0.02 & -7.58$\pm$0.02 \\
9  & 22.41$\pm$0.39 & 0.46$\pm$0.02 & -7.58$\pm$0.02 \\
10 & 21.57$\pm$0.35 & 0.51$\pm$0.01 & -7.54$\pm$0.01 \\
11 & 24.30$\pm$0.26 & 0.53$\pm$0.01 & -7.52$\pm$0.01 \\
12 & 21.45$\pm$0.31 & 0.62$\pm$0.02 & -7.46$\pm$0.01 \\
\enddata
\end{deluxetable}

\begin{deluxetable}{ccl}
\tablecaption{Best-fit parameters \label{tab:synpar}}
\tablewidth{0pt}
\tablehead{
\colhead{Model} & \colhead{Parameter} & \colhead{Best Fit}}
\startdata
Circular & $r_{\rm in}$  & 0.048 $\pm$ 0.001~au \\ 
 & $r_{\rm out}$ & 0.898 $\pm$ 0.001~au \\ 
 & $T_0$         & 2770 $\pm$ 150~K \\
 & $\alpha$      & -0.33 $\pm$ 0.05 \\
 & $N_0$         & 9.0 $\pm $2.0 $\times$10$^{19}$ cm$^{-2}$ \\
 & $\beta$       & -2.0 $\pm$ 0.1 \\
\hline
Multi-component & $r_{\rm in}$  & 0.052 $\pm$ 0.002 au \\
 & $r_{\rm out}$ & 1.53 $\pm$ 0.16 au \\
 & $r_{\rm b}$   & 0.14 $\pm$ 0.08 au \\
 & $\omega_{in}$ & $<$ 40\degr \\
 & $e_{\rm in}$  & 0.056 $\pm$ 0.015 \\
 & $e_{\rm out}$ & 0.048 $\pm$ 0.008 \\
 & $\Delta$     & 0.009 $\pm$ 0.003~au \\
 & $\iota_{\rm out}$ & 3.11 $\pm$ 0.04 \\
\enddata
\tablecomments{See Sec. \ref{sec:line_fitting} and Sec. \ref{sec:multicomponent} for descriptions of the circular disk model and the multi-component disk model, respectively.}
\end{deluxetable}

\end{document}